# Spectroscopic characterizations of the mixed Langmuir-Blodgett (LB) films of 2, 2′-biquinoline molecules: evidence of dimer formation

S. Deb, S. Biswas, S. A. Hussain and D. Bhattacharjee[*]

Department of Physics,

Tripura University,

Suryamaninagar-799130,

Tripura,

INDIA

[*] Corresponding author

Tel. +91-381-2375317

Fax: +91-381-2374801

e-mail: tuphysic@sancharnet.in

**Key words:**

Nonamphiphilie, Langmuir-Blodgett films, UV-Vis absorption and fluorescence spectroscopy.

**Abstract:**

This communication reports the $\pi$-A isotherms and spectroscopic characterizations of mixed Langmuir and Langmuir-Blodgett (LB) films of nonamphiphilic 2, 2′-biquinoline (BQ) molecules, mixed with polymethyl methacrylate (PMMA) and stearic acid (SA). The $\pi$-A isotherms and molefraction versus area per molecule studies indicate complete immiscibility of sample (BQ) and matrix (PMMA or SA) molecules. This immiscibility may lead to the formation of microcrystalline aggregates of BQ molecules in the mixed LB films. The scanning electron micrograph gives the visual evidence of microcrystalline aggregates of BQ molecules in the mixed LB films. UV-Vis absorption, fluorescence and excitation spectroscopic studies reveal the nature of these microcrystalline aggregates. LB films lifted at higher surface pressure indicate the formation of dimer or higher order n-mers.

**Introduction:**

2, 2′-Biquinoline (BQ) and its derivatives are widely used in various thin film application devices owing to their strong photoluminescence and electroluminescence properties [1]. Biquinoline derivatives have been used in thin film antenna system to mimic photosynthetic light harvesting [2]. In thin film light emitting diode some of the highly fluorescent biquinoline derivatives are used [3, 4]. When biquinoline derivatives are attached to certain other suitable liquid crystal, electro-optical display is observed [5]. Moreover 2, 2′-Biquinoline (abbreviated as BQ) is widely used as the parent compound to make drugs especially anti-malarial medicines, fungicides, biocides, alkaloids, dyes [6].

Study of organized molecular assemblies of mother chromophore 2, 2′-biquinoline (BQ) in restricted geometry of Langmuir-Blodgett film is important since in the organized LB films various physical characteristics may be manipulated with ease by changing various parameters that may lead to various technical applications.

BQ, a nonamphiphilic molecule cannot form stable LB film. However, when mixed with a long chain fatty acid (stearic acid) or an inert polymer (polymethyl methacrylate) forms stable LB films.

Here we report the detailed spectroscopic characterizations of nonamphiphilic BQ with two different supporting matrices, polymethyl methacrylate (PMMA) and stearic acid (SA) in the light of UV-Vis absorption and steady state fluorescence spectroscopic study. The most interesting feature in our observation is that BQ-SA mixed LB films are very sensitive to the pressure effect.

**Experimental**:

BQ (98% pure) purchased from Aldrich chemical, USA was purified by repeated recrystalization before use. Stearic acid (SA), purity >99%, obtained from Sigma chemical Co. and isotactic polymethyl methacrylate (PMMA) from polyscience were used as received.

The purity of the sample was checked by absorption and fluorescence spectroscopy prior to use. Spectroscopic grade ethanol (E. Merck, Germany) and chloroform (SRL, India) were used as solvents and their purity were also checked by fluorescence spectroscopy before use. Measurement of surface pressure versus area per molecule isotherms and the deposition of multilayer LB films were carried out by Langmuir-Blodgett (LB) film deposition instrument (APEX-2000C, India). The detail experimental procedure was described elsewhere [7]. All the films were 10 bilayered and the transfer ratio was found to be 0.98±0.02. Triple distilled deionised water was used as subphase and the temperature was kept at $24^0$C.

Fluorescence spectra and the UV-Vis absorption spectra of the LB films of BQ in SA as well as also in PMMA were studied by a Perkin Elmer LS-55 spectrophotometer and a Perkin Elmer Lamda-25 spectrophotometer respectively. Scanning electron micrographs was taken in a Hitachi (Japan) scanning electron microscope (model S-415A).

**Results and discussions:**

**Isotherm characteristics at the air water interface**: -

Figures 1(a) and 1(b) show the surface pressure ($\pi$) vs. area per molecule (A) isotherms of BQ mixed with PMMA and SA respectively at different molefractions of BQ alongwith pure PMMA and SA.

The area per molecule for pure PMMA and SA are 0.11 $nm^2$ and 0.23 $nm^2$ respectively at a surface pressure of 15 mN/m. At 25 mN/m the area per molecule for pure SA is 0.21 $nm^2$, which is consistent with the reported result [8]. It is also evident that in both the cases the area per molecule gradually decreases with increasing molefractions of BQ in PMMA or SA matrix.

In pure PMMA isotherm, there consist certain distinct parts [9]. The 'transition' observed at about 8 mN/m is characteristic for the isotherm of PMMA. The isotherm of PMMA also shows an inflection point at about 20 mN/m, but above this surface pressure, the monolayer no longer remains stable. Also from the isotherms of the mixed monolayer, it is observed that with increasing mole fraction of BQ, the inflection region gradually losses its distinction and at and above 0.6 mole fraction of BQ, the mixed isotherms show steep rising upto about 50 mN/m surface pressure with out any transition point.

It is also interesting to note that at higher surface pressure, in PMMA matrix, the isotherms of all molefractions of BQ almost coincide. It happens because at higher surface pressure the monolayer, which consists of PMMA matrix, is no longer remain stable [9] and actually collapsing of monolayer occurs and multilayer and three-dimensional crystalline aggregates are formed at the air-water interface. Thus at higher surface pressure all the isotherms of mixed monolayers in PMMA matrix lose their distinguishability and overlap on each other.

From the isotherm characteristics of the mixed BQ-SA films at various molefractions of BQ (figure (1b)), it is observed that with increasing molefractions of BQ, the mixed isotherms of SA-BQ decrease systematically and even at higher molefraction, they don't coincide.

It may be mentioned in this context that the thermodynamic nature of mixing of various two-component systems was established by analyzing the $\pi$-A isotherms of the pure components as well as the binary mixtures [10]. The excess areas of mixing $A^E$ which provides a measure of non-ideality, were calculated as a function of surface pressure, mixture of composition and the molecular weight using the additivity rule [11-13]

$A^E = A_{12} - A_{av}$ ………………..(1)

$A_{av} = N_1 A_1 + N_2 A_2$ ……………(2)

Where $A_{12}$ is the actual area per molecule of mixed monolayer, $A_{av}$ is the calculated average area per molecule of mixture, assuming molefraction additivity, $A_1$ and $A_2$ are the area per molecule (or monomer) of each of the single component at a specific surface pressure, $N_1$ and $N_2$ are their corresponding molefractions.

The insets of figures 1(a) and 1(b) show the plot of the data of the actual area ($A_{12}$) per molecule versus molefraction of BQ in the mixed monolayer with PMMA and SA respectively, at the air-water interface and at a surface pressure of 15mN/m. The solid line in the figure represents the ideality curve that corresponds to the plot of average area per molecule ($A_{av}$) of the mixture versus molefraction, assuming molefraction additivity.

The experimental data in case of both PMMA and SA matrices almost coincide with the ideality curve. That is from equation (1) and (2) we may say that $A^E = 0$ and

$$A_{12} = A_{av} = N_1 A_1 + N_2 A_2.$$

It should be mentioned in this context that for binary mixture of molecules exhibiting either ideal miscibility or complete immiscibilty, $A^E = 0$, which is our case. It is important to note that in bulk binary miscible mixtures, if two liquids are ideally miscible, it usually means that they are physically and chemically similar, so that the intermolecular forces of these two components are identical and the presence of the solute does not affect the solvent molecules. However, complete immiscibilty means strong attractive interactions among like molecules and almost no interaction between unlike molecules [10]. The existence of strong attractive interaction among BQ molecules may lead to the formation of clusters and aggregates.

The plot of the collapse pressure versus molefractions of BQ in the mixed films of BQ with PMMA and SA are also shown in the insets of figures 1(a) and 1(b) respectively. It is evident from the figures that the collapse pressure for all the molefraction are almost constant and independent of molefractions of BQ in both the matrices and do not match with the ideality (solid line) curve, which indicate that BQ molecules and PMMA/SA molecules are totally immiscible in the mixed monolayer. This immiscibility or complete demixing between BQ and PMMA or SA may lead to the formation of aggregates of BQ molecules in the mixed films, which was later confirmed by Scanning electron microscopy (SEM). The nature of these aggregates have been revealed by UV-Vis absorption and steady state fluorescence spectroscopic study.

**Scanning electron microscopy:**

To confirm the formation of microcrystalline aggregates of BQ molecules in the mixed LB films we have employed traditional structural studies namely, Scanning electron microscopy (SEM). Figure 2 shows a scanning electron micrograph of the mixed LB film of 0.5M of BQ in SA matrix. The aggregates with sharp and distinct edges correspond to the three dimensional aggregates of BQ in LB films. The smooth background corresponds to the SA matrix. The formation of distinct crystalline domains of BQ, as evidenced from the SEM, provides compelling visual evidence of aggregation of BQ in the LB films.

**Spectroscopic characterizations**:

Figures 3(a) and 3(b) show the UV-Vis absorption and steady state fluorescence spectra of the mixed LB films of BQ in PMMA and SA matrices respectively at different molefractions of BQ along with the solution and microcrystal spectra for comparison. Also figures 3(a) and 3(b) show solution spectra of two molefractions of BQ (0.1 M and 0.8 M) mixed with PMMA and SA respectively.

Pure BQ solution absorption spectrum shows distinct band positions in 225-350 nm region having 0-0 band at around 340 nm and a prominent high energy band at 261 nm along with the vibrational band at around 329 nm. Also there is a weak hump at around 237 nm. Solution absorption spectra of the two molefractions of BQ (0.1M & 0.8M), mixed with PMMA and SA, give almost identical spectra as that of pure solution except that a prominent high-energy band is observed at around 233 nm. This high-energy band is quite prominent and blue shifted in the LB films and in the microcrystal spectrum.

The absorption spectra of the mixed LB films at different molefractions of BQ in PMMA and SA matrices are almost identical and have two prominent high energy bands with peak at around 255 nm and 216 nm. These two bands may originate due to blue shifting of 261 nm band and 237 nm weak hump in solution spectrum. Moreover LB films absorption spectra also show a broad low intense longer wavelength band with peak at around 330 nm. Microcrystal spectrum also has distinct similarity with LB films spectra, although little variation in band positions and intensity distribution patterns are observed.

The close similarity of the absorption spectra of the mixed LB films with that of microcrystal spectrum may lead to the conclusion that at least low dimensional microcrystalline aggregates are formed in the mixed LB films. This is also confirmed by Scanning electron micrograph. The nature of formation of these microcrystalline aggregates has been revealed by the fluorescence spectroscopic study.

There are certain rigid nearly planar plate like molecular structure namely, pyrene [14] or rod like molecules as anthracene [15] form aggregate in the mixed LB films. These molecules are actually sandwiched among the matrix molecules (namely, stearic acid) to form aggregates in the LB films. BQ molecule has almost linear rod like structure. For aggregation to be occurred in the mixed LB films of BQ molecules, the most possibility is the sandwich of BQ molecules within SA or PMMA molecules in the LB films.

The fluorescence spectrum of pure BQ solution shows a broad and intense band profile in the 325-450 nm region. This broad band actually consists of several diffused overlapping vibrational bands with 0-0 band at around 368 nm. The large shift of 0-0 band in fluorescence spectrum in comparison to the solution absorption spectrum may be due to the deformation produced in the electronic states of the molecule. The solution spectra of BQ mixed with PMMA/SA in two different molefractions (0.1M and 0.8M of BQ) are totally identical with that of pure BQ solution spectrum. This may lead to the conclusion that no change in molecular electronic structure is occurred in the mixed solution.

The microcrystal fluorescence spectrum shows prominent and intense 0-0 band at 396 nm alongwith the weak hump at 441 nm and an almost indistinguishable hump at around 417 nm which is intense and prominent in the mixed LB films. The high-energy band at 368 nm in solution spectrum is totally absent in microcrystal spectrum. The total quenching of this high-energy band is due to the reabsorption effect arising due to the formation of microcrystalline aggregates and have been observed in several other molecules [16]. The fluorescence spectra of the mixed LB films of BQ in PMMA and SA matrices at different molefractions of BQ also have distinct similarity with the microcrystal spectrum with 0-0 band at 396 nm. However, certain distinct features are observed in the mixed LB films. Especially at low molefractions of BQ in PMMA matrix (figure 3(a)) unlike that of microcrystal spectrum, distinct vibrational pattern is observed in the longer wavelength region with 0-0 band at 396 nm, intense peak at 417 nm and a weak hump at around 441 nm. With increasing molefractions, longer wavelength 417 nm band is gradually reduced in intensity and high-energy 0-0 band at 396 nm becomes intense and at higher molefraction of 0.8 M the LB films fluorescence spectrum is almost similar to that of the microcrystal spectrum. Mixed LB films of BQ in SA matrix at various molefractions also exhibit similar band pattern. However, here even at very low molefractions, 417 nm band is comparatively low intense than that of 396 nm band unlike that in PMMA matrix. This may be attributed solely to the different microenvironment in PMMA and SA matrices. PMMA as a coil like molecule [9] may create a microcavity where BQ molecules are inserted. Where as SA molecules are rod like and BQ molecules may be sandwiched among SA molecules.

The longer wavelength distinct vibrational pattern in the mixed LB films is neither observed in microcrystal spectrum nor in solution spectrum. This may arise because in the process of fabrication of LB films various parameters may play an important role and here in the case of mixed LB films molecular electronic states of BQ may be considerably deformed.

This may happen because there is possibility of BQ molecules become twisted in the microenvironment of PMMA/SA matrices.

It should be mentioned in this context that terphenyl [16] molecules are planar in crystal and LB films at room temperature but are twisted in solution phase. Due to the change in molecular structure terphenyl molecules in the mixed LB films and in crystal give highly quenched high-energy band in the fluorescence spectrum. However, unlike terphenyl molecule here in the mixed LB films of BQ molecules, the high-energy band at 396 nm becomes gradually intense with increasing molefractions of BQ. In the microcrystal spectrum high-energy band at 396 nm becomes highly intense. This may be due to the fact that at lower molefractions in the mixed LB films due to the presence of large amount of PMMA/SA molecules, BQ molecules may be twisted more so that the degree of deformation in the molecular electronic states is appreciably large. With increasing molefractions of BQ in the mixed LB films, degree of deformation tends to reduce due to the lack of larger amount of matrix molecules.

**Layer effect:**

Various technical applications may sometimes require thick films. Figures 4(a) and 4(b) show the UV-Vis absorption and fluorescence spectra of different layered mixed LB films of BQ in PMMA and SA matrices respectively at two different molefractions of 0.1 M and 0.5 M of BQ.

The UV-Vis absorption spectra of the mixed LB films in both the matrices show almost similar band pattern except a small change in intensity distribution. However, their fluorescence spectra are somewhat different. The fluorescence spectra of less number of layers with low molefraction of BQ the mixed LB films of both the matrices show well structured vibrational pattern with intense low energy vibrational band at 417 nm and comparatively low intense high energy band at 396 nm. Especially in PMMA matrices 417 nm band is comparatively intense with respect to 396 nm band at 0.1M of BQ in 5-layer LB film unlike that in SA matrix. This may be due to the different kinds of microenvironments into different matrices and has been discussed before. However, with increase in number of layers and also at higher molefraction, high energy band at 396 nm becomes intense where as longer wave length band reduces to a weak hump. Therefore it may be concluded that with increasing thickness of LB films and also at higher molefractions, closer association of molecular organization of BQ molecules occur resulting in the decrease of deformation of molecular electronic level.

**Pressure effect:**

The morphology and the crystal parameter of the LB films may be controlled by varying the surface pressure of lifting. Here we have studied the spectroscopic characteristics of mixed LB films of BQ in SA at two different molefractions (0.1 M and 0.5 M of BQ) using 15, 20, 25, 30 mN/m surface pressure. All the films are 10 bilayered.

For lower molefraction of 0.1 M of BQ in SA, no appreciable change is observed in both absorption and fluorescence spectra with change in pressure. However, the absorption spectra of the mixed LB films of BQ at a higher molefraction of 0.5 M show interesting change with the increase in surface pressure as shown in figure 5(a). The UV-Vis absorption spectra of mixed LB films of BQ lifted at high surface pressure of 20 and 30 mN/m are of totally different in nature. Unlike the previous spectra here the most interesting thing is that a longer wavelength broad and low intense band with peak at around 350nm and a kink at around 385 nm are observed. Moreover the higher energy bands are also somewhat diffused and the absorption spectra are totally different in nature than that lifted at lower surface pressure and lower molefraction. In the fluorescence spectra also only high energy 0-0 band at around 396 nm become prominent and other low energy bands are reduced to weak hump. Although the nature of fluorescence spectra are explained before however the over all different nature in the absorption spectra at higher surface pressure cannot be explained readily. One possibility may be due to the formation of dimer or higher order n-mers.

The excitation spectra of the mixed LB films of 0.5M of BQ in SA matrix lifted at different surface pressure along with the microcrystal spectrum are shown in figure 5(b). The monitoring wavelength is 417 nm. It is interesting to note that although the excitation spectra of mixed LB films lifted at 15 mN/m surface pressure closely resembles to that of microcrystal spectrum. Where as the excitation spectra of the Mixed LB films of higher surface pressure namely, 20, 25 and 30 mN/m surface pressure, are totally different in nature. This firmly corroborates the fact that the origin of the excitation spectra of LB films lifted at lower surface pressure and higher surface pressure are from totally different kind of species. Especially at high surface pressure the excitation spectra may originate from dimeric or higher order n-meric species.

**Conclusion:**

In conclusion, our results show that non-amphiphilic 2, 2′-Biquinoline forms an excellent Langmuir monolayer at the air-water interface when incorporated into PMMA or SA matrices and can be easily transferred onto quartz substrates to form mono and multilayer LB films. Isotherms as well as area per molecule versus molefraction studies of BQ reveal either ideal mixing or complete demixing of BQ molecules with the matrix molecules. However the plot of collapse pressure versus molefractions of BQ in the mixed LB films indicate that BQ molecules and PMMA or SA molecules are immiscible in the mixed monolayer, which leads to the formation of aggregates of BQ molecules in the mixed LB films. Scanning electron micrograph gives visual evidence of microcrystalline aggregates of BQ molecules in the mixed LB films. UV-Vis absorption and steady state fluorescence spectroscopic study reveal the nature of these microcrystalline aggregates in LB films. The fluorescence spectra of BQ in mixed LB films and in microcrystal are quite different from that in ethanol solution and have been explained to be due to the reduction of degree of deformation of the electronic states when BQ molecules goes from solution to solid states/films. Dimer and higher order n-mers are formed in the mixed LB films of higher molefraction of BQ in SA matrix, which is lifted at higher surface pressures of 20, 25 and 30 mN/m.

**References:**

1. F. Liang, J. chen, Y. Cheng, L. Wang, D. Ma, X. Jing, F. Wang, J. Mater. Chem. 13 (2003) 1392.

2. H. B. Baudin, J. Davidson, S. Serroni, A. Juris, V. Balzani, S. Canpagna, L. Hammarstron, J. Phy. Chem A. 106 (2002) 4312.

3. T. Mori, K. Obata, T. Mizutani, J. Phys. D: Appl. Phys. 32 (1999) 1198.

4. T. Mori, H. Tsuge, T. Mizutani, J. Phys. D: Appl. Phys. 32 (1999) L65.

5. D. Schneider, T. Rabe, T. Riedl, T. Dobbertin, M. Kroger, E. Becker, H. H. Johannes, W. Kowalsky, T. Weimann, J. Wang, P. Hinze, Appl. Phys. Lett. 85 (2004) 1886

6. Toxicological Review of Quinoline (CAS No. 91-22-5), U.S. Environmental Protection Agency, Washington, DC, 2001.

7. S. Acharya, D. Bhattacharjee, G.B. Talapatra, Chem. Phys. Lett. 352 (2002) 317.

8. K. Ray, D. Bhattacharjee, T. N. Misra, J.Chem. Soc. Farady Trans. 93 (1997) 4041.

9. R. H. G. Brinkhuis, A. J. Scchouten, Macromolecules 24 (1991) 1487.

10. N. Sadrazedh, H. Yu, G. Zografi, Langmuir 14 (1998) 151.

11. Gaines, G. L, Ed. Insoluble Monolayers at Liquid-Gas Interfaces Wiley Interscience, New York, 1966.

12. Dorfler, H. D. Adv. Colloid Interface Sci. 31 (1990) 1.

13. Barnes, G.T., J. Colloid Interface Sci. 144 (1991) 299.

14. A. K. Dutta, T. N. Mishra, A. J. Pal, Langmuir 12 (1996) 359.

15. A. K. Dutta, Langmuir 13 (1997) 5678.

16. A. K. Dutta, T. N. Mishra, A. J. Pal, J. Phys. Chem. 98 (1994) 12844.

**Figure captions:**

Fig. 1 Surface pressure (π) versus area per molecule (A) isotherms of mixed Langmuir films of BQ in (a) PMMA matrix (b) SA matrix. The numbers denote the corresponding mole-fractions of BQ in PMMA/SA matrix. The insets show plots of collapse pressure versus molefractions of BQ and area per molecule versus mole-fractions of BQ at 15 mN/m surface pressure. Molecular structure of BQ is also shown in the inset of figure (a).

Fig. 2. SEM of 10 layers of BQ-SA mixed LB films (molefraction of BQ 0.5M) at room temperature.

Fig. 3. UV-Vis absorption and steady state fluorescence spectra of mixed LB films of BQ in (a) PMMA matrix and (b) SA matrix, along with solution (EtOH) and microcrystal (MC) spectra. The numbers denote the corresponding mole-fractions of BQ in PMMA/SA matrix.

Fig. 4. UV-Vis absorption and steady state fluorescence spectra of mixed LB films of BQ in (a) PMMA matrix and (b) SA matrix for different layers at two mole-fractions of 0.1 M and 0.5 M of BQ.

Fig. 5(a) UV-Vis absorption and steady state fluorescence spectra of BQ/SA LB films lifted at different surface pressure at two different molefractions of 0.1M and 0.5 M of BQ. (b) Excitation spectra of BQ/SA LB films and microcrystal lifted at different surface pressure. The monitoring wavelength is 418 nm. The numbers denote the corresponding surface pressure of lifting.

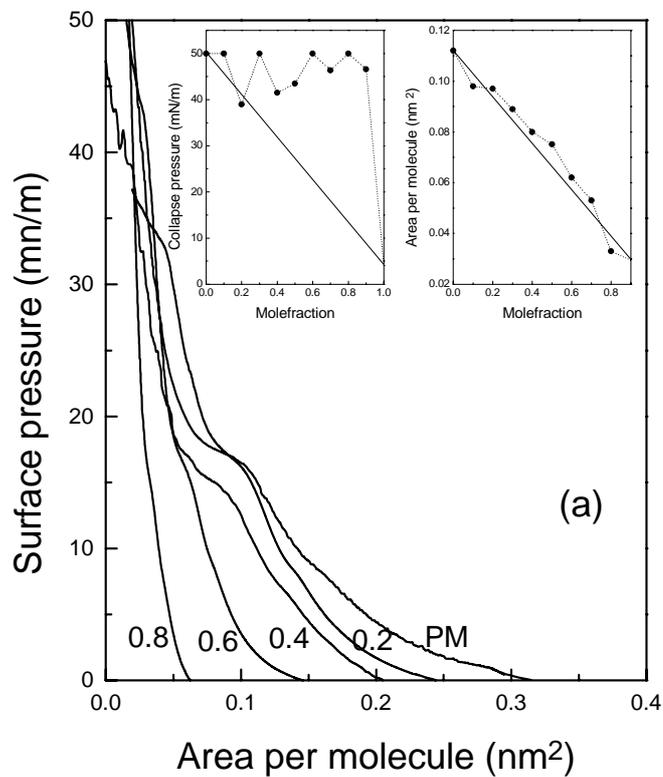

Figure: 1(a) S. Deb et. al

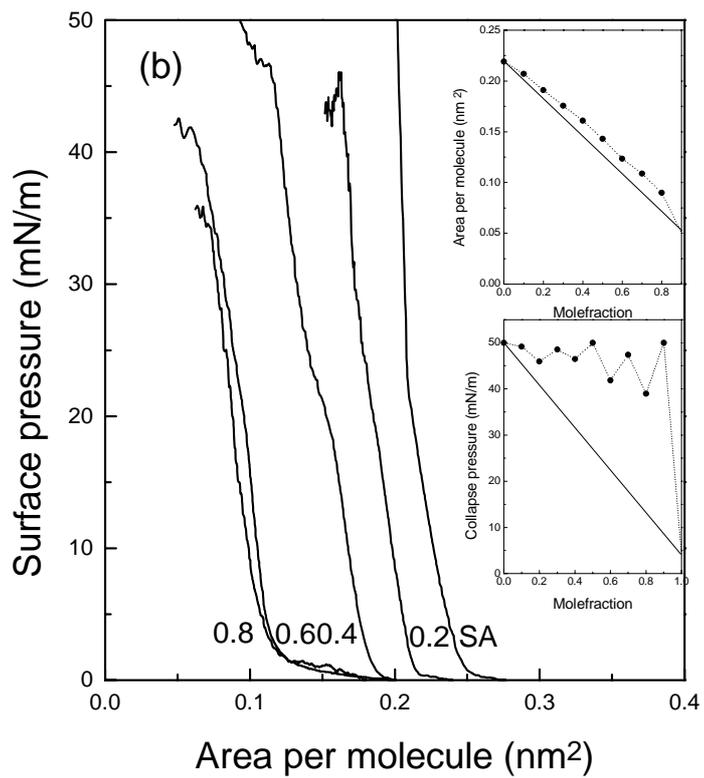

Figure: 1(b) S. Deb et. al

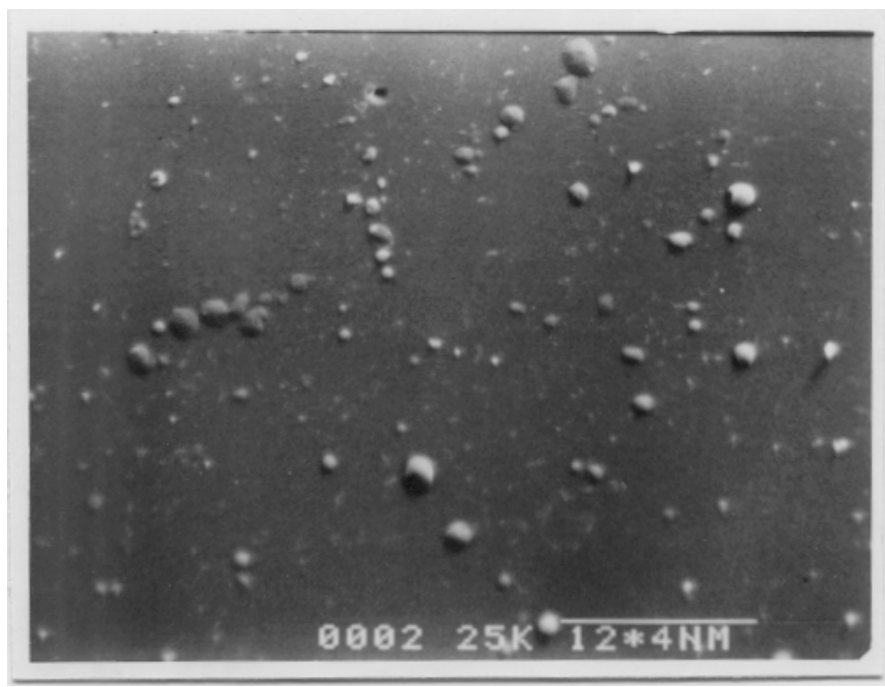

Figure: 2. S. Deb et.al.

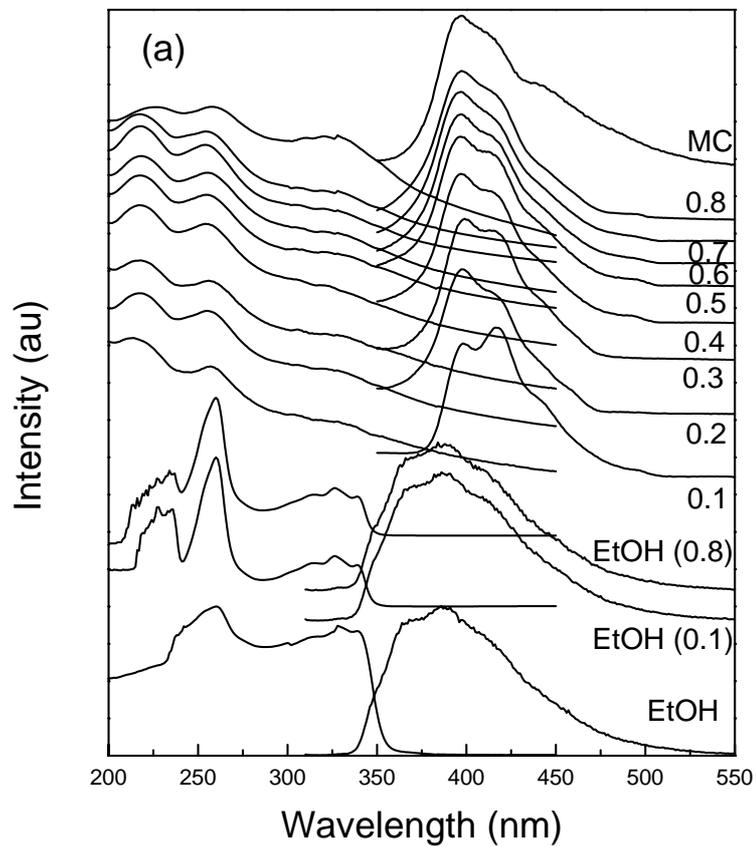

Figure: 3(a) S. Deb et. al

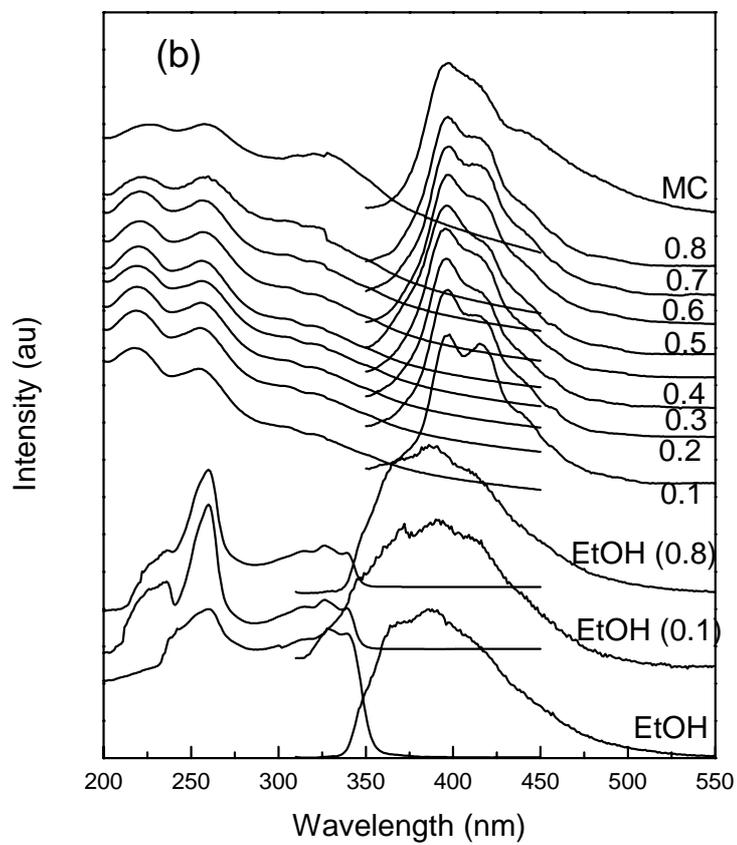

Figure: 3(b) S. Deb et. al

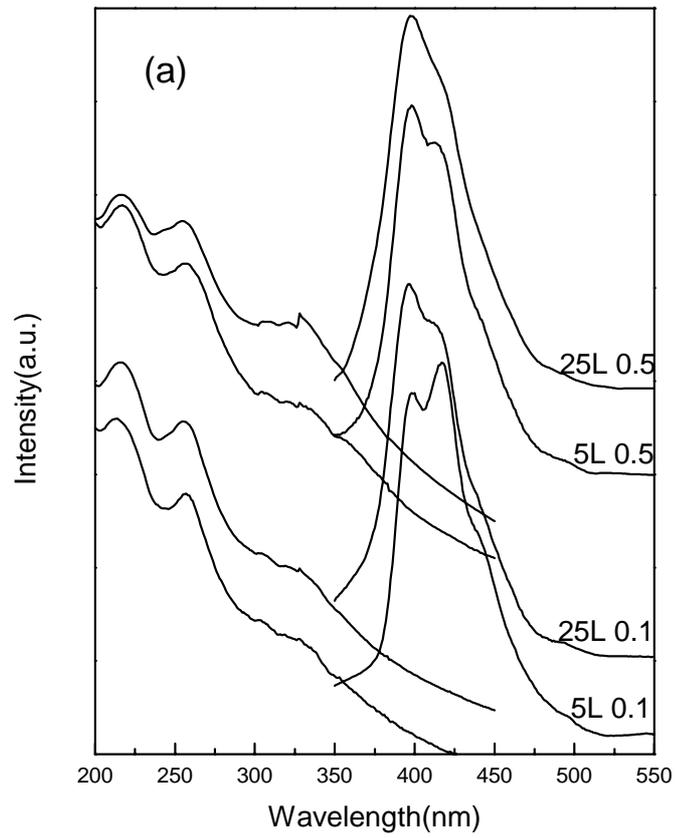

Figure: 4(a) S. Deb et. al

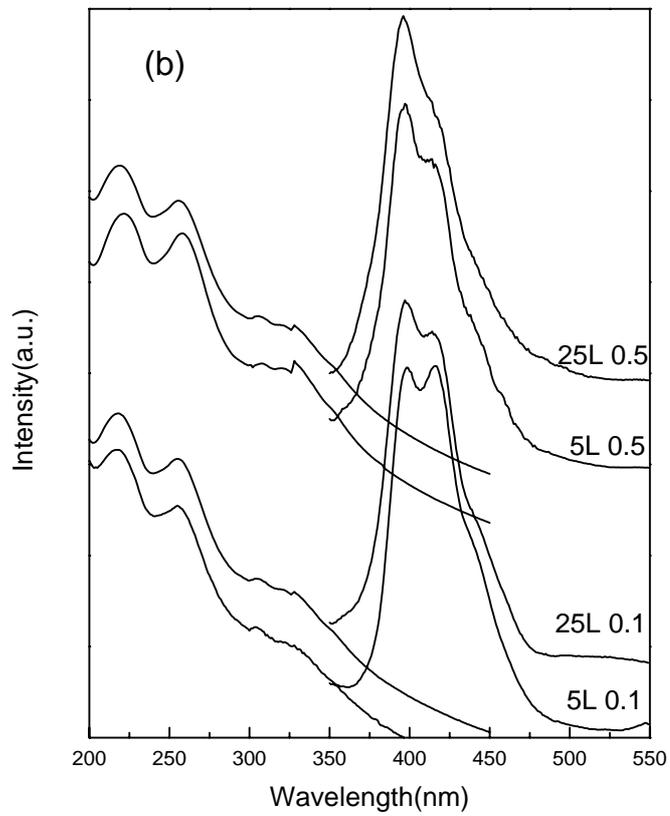

Figure: 4(b) S. Deb et. al

21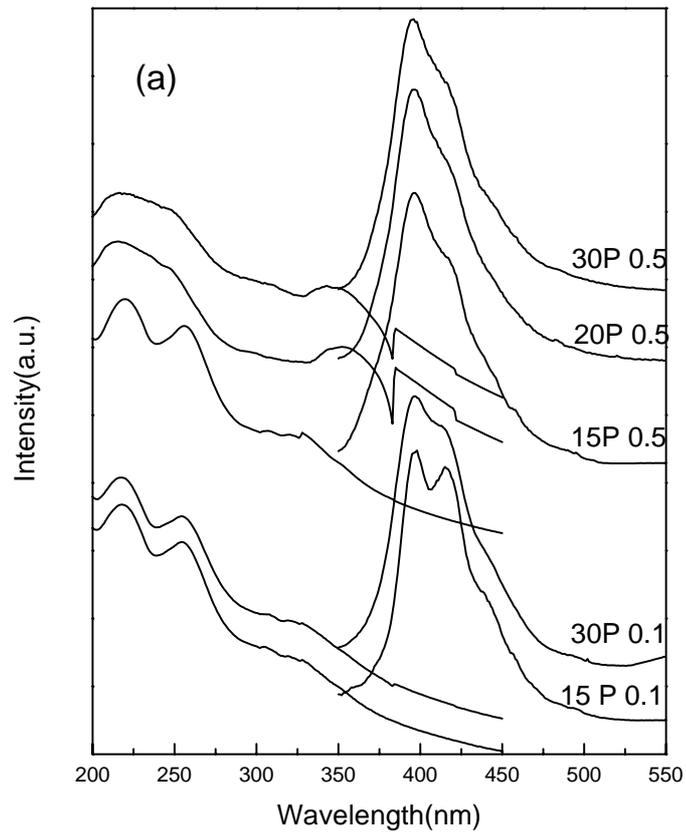

Figure: 5(a) S. Deb et. al

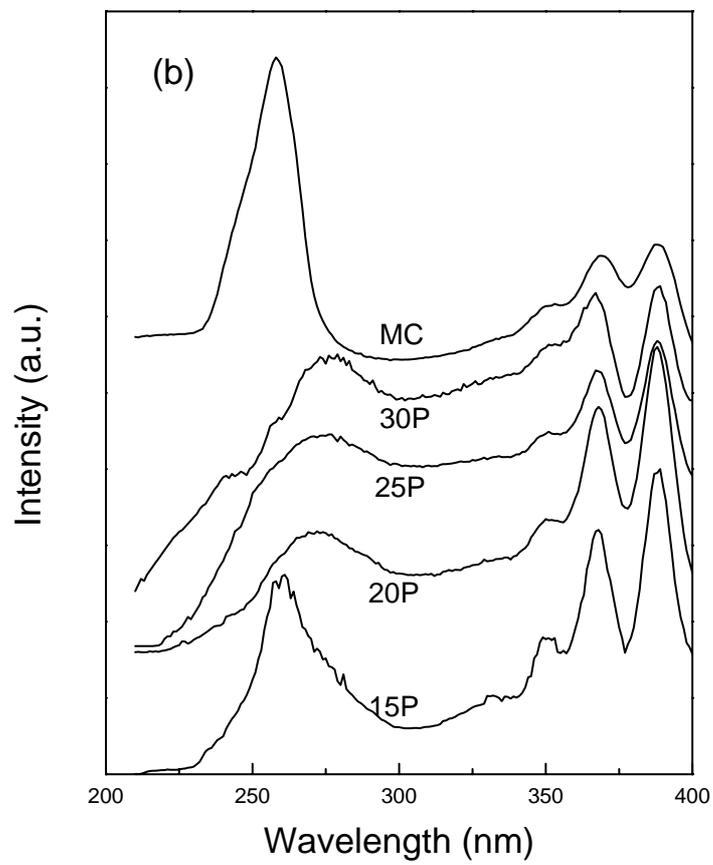

Figure: 5(b) S. Deb et. al